\documentstyle[11pt,paspconf,epsf]{article}

\newcommand{\ltsima} {$\; \buildrel < \over \sim \;$}
\newcommand{\gtsima} {$\; \buildrel > \over \sim \;$}
\newcommand{\lta} {\lower.5ex\hbox{\ltsima}}
\newcommand{\gta} {\lower.5ex\hbox{\gtsima}}

\begin{document}

\title{THE INTRINSIC UV/SOFT X-RAY SPECTRUM OF QUASARS}

\author{Francesco Haardt}
\affil{Department of Astronomy \& Astrophysics, Institute of
Theoretical Physics, G\"oteborg University \& Chalmers University
of Technology, 412 96 G\"oteborg, Sweden}

\author{Piero Madau}
\affil{Space Telescope Science Institute, 3700 San Martin
Drive, Baltimore MD 21218, USA}

\begin{abstract}

The detection of a HeII  absorption trough in the spectra of three
high-redshift quasars provides new constraints on the spectral shape of the UV
extragalactic background. The relative strengths of the observed flux decrements
at the rest-frame wavelengths of 1216 and 304 \AA\ require a relatively soft
radiation field at 4 ryd compared to 1 ryd. If the ionizing metagalactic flux
is dominated by the integrated light from QSOs, and the HeII 
reionization of 
the universe was completed well before $z\sim 3$, then the UV/soft X-ray
spectrum of individual quasars at these epochs must, once the cosmological
``filtering'' through material along the line of sight is taken into account,
satisfy similar constraints on the average. We model the propagation of
AGN-like ionizing radiation through the intergalactic medium using CUBA, a
numerical code developed in our prior work. We show that, in order to explain
the reported HeII  absorption, any thermal component responsible for the ``soft
X-ray excess'' observed in the spectra of AGNs at $z\lta 0.3$ must, at
$z\sim 3$, have a typical temperature $\gta 80$~eV, and a luminosity not
exceeding $\approx 20$\% of that of the ``UV bump''. 

\end{abstract}


\section{Results}

By means of the code named CUBA (described in HMI), we have 
calculated the ratio $\eta\equiv N_{\rm HeII}/N_{\rm HI}$ 
for a set of QSO spectral models as discussed in HMII. 

The HI and HeII  
opacities computed with various QSO background models have been then
tested against the observed values. Together with the already mentioned 
HeII 
data, we have taken a collection of HI data. The calculated redshift
dependent HeII $D_A$ (the fractional flux decrement shortward of Ly$\alpha$)  
is shown in Figure for different input QSO spectral models,
together with the available data points. In the left upper panel we plotted
also HI $D_A$ against the data. 

From the Figure it is evident that:

$\bullet$ The spectral slope below $\sim 1000$\AA ~of integrated QSO spectrum 
must be comprised between 1.5 and 2.

$\bullet$ The soft excess observed in QSO spectra at $z\lta 0.3$ must 
have at $z\sim 3$ a temperature $\gta 80$ eV, and a luminosity not 
exceeding $\approx 20$\% of that of the UV bump.

A detailed analysis and discussion, with all the relevant references, will 
appear in HMII.

\begin{figure}
\plotone{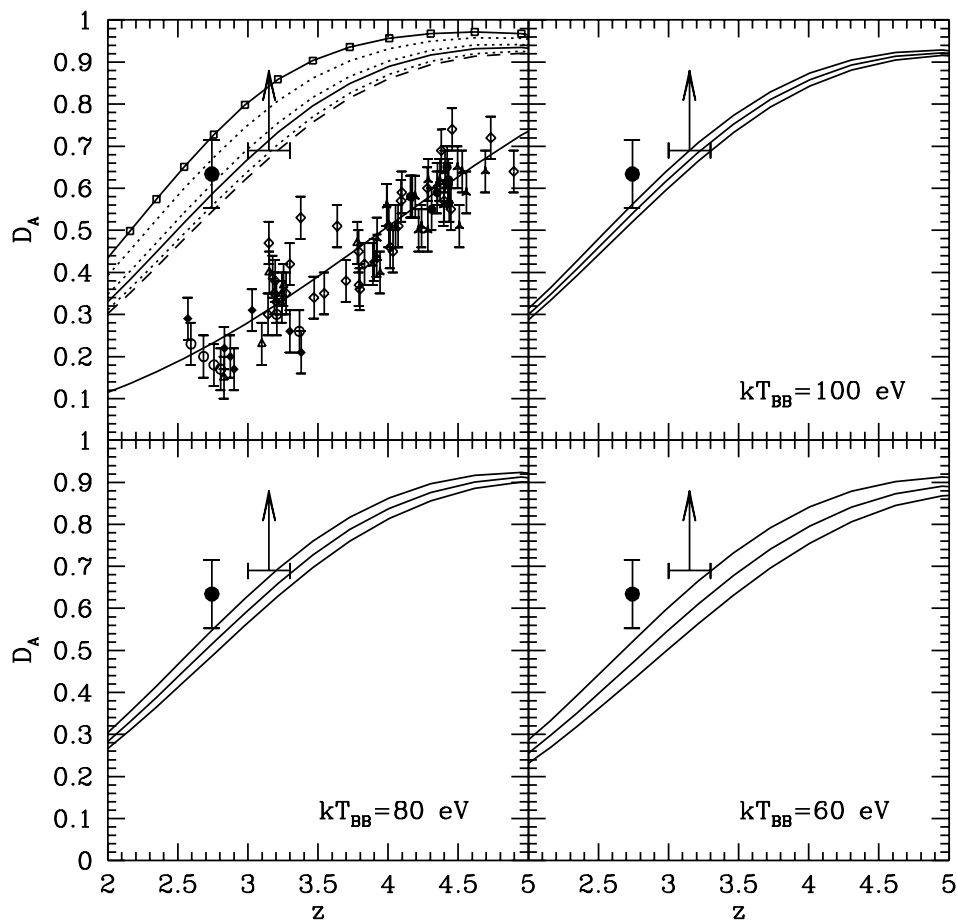}
\caption{The quantity $D_A$ for HI and HeII computed for various 
QSO spectral models is tested against existing data.}
\end{figure}


\begin{thebibliography}{}

\bibitem[]{} Haardt F., Madau P., 1996, \apj, 461, 20 (HMI)

\bibitem[]{} Haardt F., Madau P., 1996, in prep. (HMII)

\end{thebibliography}
\end{document}